%% file: main.tex
\documentclass[10pt,twocolumn,letterpaper]{article}

\usepackage[pagenumbers]{cvpr} 

\input{preamble}

\usepackage{capt-of}

\definecolor{cvprblue}{rgb}{0.21,0.49,0.74}
\usepackage[
    pagebackref,
    breaklinks,
    colorlinks,
    allcolors=cvprblue
]{hyperref}

\def\paperID{*****} 
\def\confName{3DV\xspace}
\def\confYear{2027\xspace}

\title{TopoRig: Topology-Agnostic Facial Rigging via Multi-Source Supervision}

\author{
Andrew Fleet$^{1,4,5}$ \qquad
Soroush Mehraban$^{2,3,4,5}$ \qquad
Vida Adeli$^{2,3,4,5}$\\[3pt]
Cole Clifford$^{2}$ \qquad
Babak Taati$^{3,4,5}$\\[7pt]
$^{1}$Queen's University \qquad
$^{2}$Pickford AI \qquad
$^{3}$University of Toronto\\[2pt]
$^{4}$Vector Institute \qquad
$^{5}$KITE Research Institute
}
\begin{document}

\twocolumn[{%
\renewcommand\twocolumn[1][]{#1}%
\maketitle

}]

\begin{abstract}
Automatic facial rigging across heterogeneous mesh topologies remains challenging because high-quality expression supervision is often tied to canonical templates, while deformation transfer to arbitrary meshes can introduce geometric artifacts and correspondence errors. We present \textbf{TopoRig}, a topology-agnostic facial rigging framework that predicts FACS-conditioned deformations directly on input mesh vertices while preserving the original topology. Starting from the ICT FaceKit expression model, we construct complementary supervision from accurate but template-biased common-topology rigs, topology-diverse but noisier transferred rigs, and targeted image-based cues for controls poorly captured by geometric transfer. TopoRig combines local surface geometry, landmark-relative semantic features, global shape context, and FACS controls to predict per-vertex displacements. We train on 3,496 generated identities using 45 non-gaze expression controls
from the 53-control ICT FaceKit vocabulary. On held-out identities and unseen mesh topologies, TopoRig more faithfully reproduces the reference expression space than prior neural facial-rigging methods, while qualitative results show consistent localized deformations across diverse character geometries. Ablations demonstrate that semantic landmark features and complementary supervision improve cross-identity and cross-topology generalization. Overall, TopoRig amortizes heterogeneous and imperfect expression supervision into a single topology-preserving deformation model.
\end{abstract}

\input{sections/1_introduction}
\input{sections/2_related_works}
\input{sections/3_method}
\input{sections/4_experiments}
\input{sections/5_conclusion}

{
    \small
    \bibliographystyle{ieeenat_fullname}
    \bibliography{main}
}

\input{sections/appendix}

\end{document}

%% file: preamble.tex
\usepackage{booktabs}
\usepackage{multirow}

%% file: sections/1_introduction.tex
\section{Introduction}

\begin{figure}[t]
    \centering
    \includegraphics[width=\columnwidth]{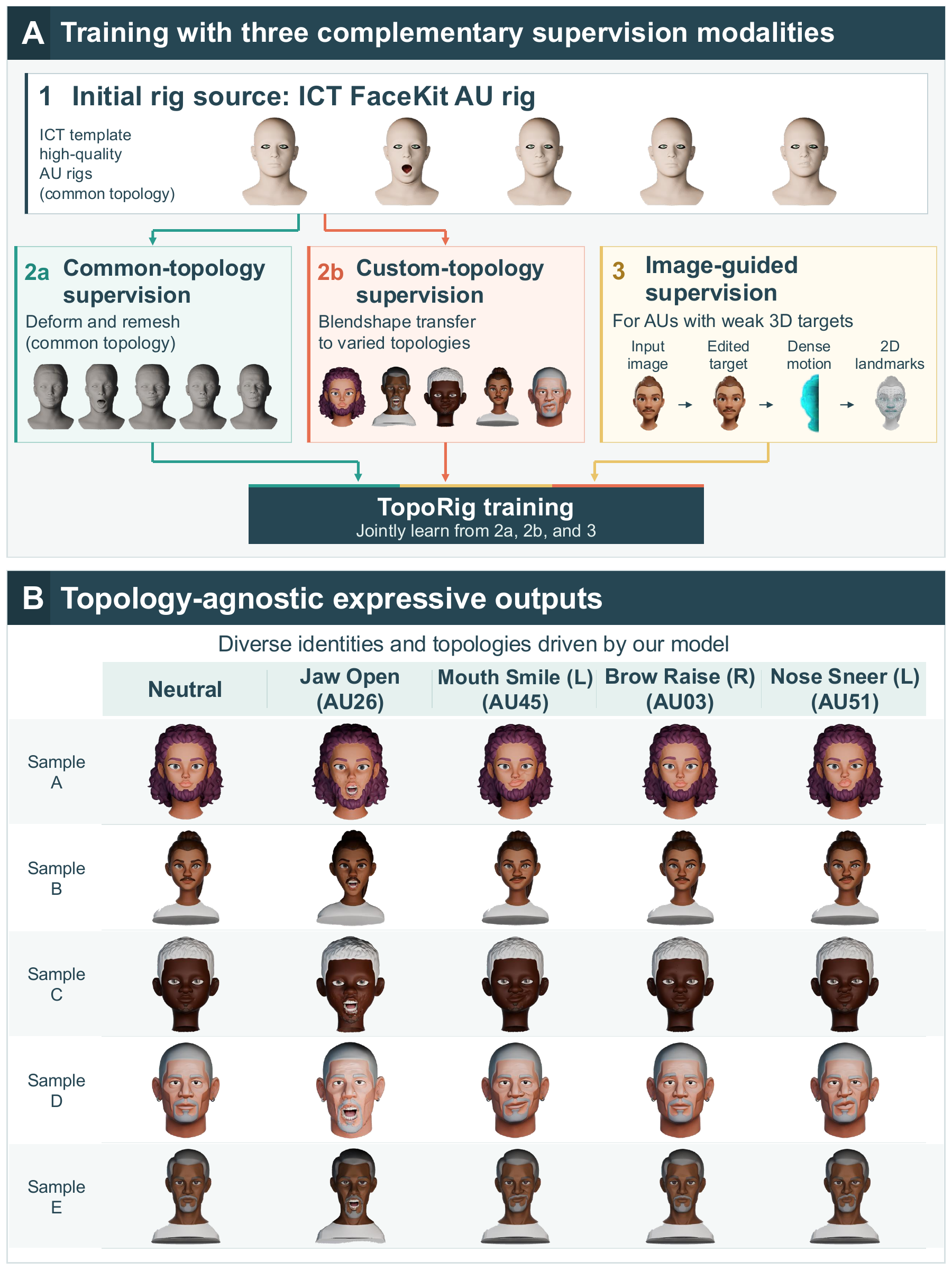}
    \caption{
    \textbf{Overview of TopoRig.}
    \textbf{(A)} Starting from the ICT FaceKit AU rig, we construct complementary supervision from common-topology fitted rigs, expression transfer to custom topologies, and targeted image-based cues for AUs with unreliable 3D targets. TopoRig jointly learns from these sources.
    \textbf{(B)} At inference time, it produces AU-conditioned deformations for diverse identities and previously unseen mesh topologies while preserving the input topology.
    }
    \label{fig:overview}
\end{figure}

Facial rigging transforms a static neutral face into a controllable asset capable of producing a range of expressions~\cite{li2010example, qin2023nfr}. It is an important component of creating digital characters for user-controlled animation, performance retargeting, generated avatars, speech-driven animation, and stylized expression manipulation~\cite{xu2014controllable,thies2016face2face,richard2021meshtalk,li2025aublendnet}. Facial rigs commonly represent expressions using blendshapes corresponding to the Facial Action Coding System (FACS) ~\cite{ekman1978facs, lewis2010direct, lewis2014practice}, providing interpretable controls that can be adjusted individually or combined to form more complex expressions. However, manually constructing high-quality facial rigs requires substantial artist effort, resulting in high production costs and long production cycles~\cite{wang2023versatile,hou2024neutral,li2017flame}.

Traditional automatic methods reduce this effort by fitting a predefined facial template and transferring its blendshapes to a target mesh, but this process often relies on template registration, dense correspondence, or topology normalization~\cite{dutreve2010easy,onizuka2019landmark,carrigan2020expression}. Recent neural approaches instead predict facial deformations directly, with some supporting expression transfer across meshes with different connectivity and other combining global expression controls with localized deformation~\cite{qin2023nfr,cha2025nfs,ma2025riganyface}. However, methods built on a canonical mesh representation, such as AU-Blendshape, remain limited to a shared topology. Template-registration approaches such as OmniFaceRig support more diverse input meshes, but construct the final facial rig through fitted template geometry and blendshape transfer~\cite{li2025aublendnet,wang2026omnifacerig}. RigAnyFace supports varied input topology and expands training through 2D supervision, enabling facial rigging across diverse mesh structures~\cite{ma2025riganyface}. 

Despite the progress, a central challenge in topology-agnostic facial rigging is obtaining supervision that is simultaneously accurate, geometrically diverse, and applicable across heterogeneous mesh connectivity. To address this challenge, we construct a multi-source training corpus using the open-source ICT FaceKit expression model~\cite{li2020learning} together with diverse neutral 3D faces generated by an off-the-shelf image-to-3D model~\cite{li2026pixal3d}. We derive 3D expression supervision through two complementary routes. First, we fit the ICT FaceKit template to the generated identities, producing accurate and internally consistent expression targets on a shared topology, although the resulting geometry remains biased toward the template. Second, we transfer the ICT FaceKit expression deformations directly onto the generated meshes, preserving their identity-specific geometry while substantially increasing topology diversity. Although the transfer procedure includes geometric smoothing and expression-specific corrections, local correspondence ambiguities can still produce imperfect targets, particularly around closely separated facial surfaces such as the eyelids and lips. Moreover, the procedure requires substantial per-identity processing. We therefore use these transferred meshes as supervision for an amortized deformation model rather than as the final rigging procedure. For action units that are poorly captured by geometric transfer, we additionally incorporate targeted image-based supervision obtained using an off-the-shelf image editing model~\cite{google2025nanobanana}.

We propose \textbf{TopoRig}, a topology-agnostic facial rigging framework that learns a single deformation model from complementary supervision sources. Rather than relying on one automatically constructed rig representation, TopoRig combines accurate but template-biased fitted rigs, topology-diverse but noisier transferred rigs, and targeted image-based supervision for controls that are difficult to obtain reliably through geometric transfer. The network conditions each input vertex on local geometry, optional landmark-relative semantic features, global facial geometry, and the requested FACS control, and directly predicts expression-dependent displacements on the original mesh vertices. This preserves the input vertex count, connectivity, and UV parameterization while amortizing the expensive per-identity fitting and transfer process into a single forward pass.

Fig.~\ref{fig:overview} summarizes the construction of the three
supervision sources and the topology-preserving inference process. In summary, our main contributions are:
\begin{itemize}
    \item \textbf{Complementary multi-source supervision:}
    a training strategy that derives complementary supervision from a common AU rig through fitted common-topology targets, deformation transfer to identity-specific topologies, and targeted image-based cues for controls with unreliable 3D targets.

    \item \textbf{Amortized topology-preserving facial deformation:}
    a FACS-conditioned deformation model that jointly learns from these heterogeneous supervision sources and predicts expressions directly on previously unseen input meshes without running the fitting and transfer pipeline at inference time.

    \item \textbf{Large-scale training and analysis:} a training corpus
    containing 3,496 generated identities, using 45 non-gaze controls
    from the 53-control ICT FaceKit vocabulary, together with evaluations
    and ablations studying cross-topology generalization, semantic landmark
    features, complementary training sources, and data scale.
\end{itemize}

%% file: sections/2_related_works.tex
\section{Related Work}

\begin{figure*}[!t]
    \centering
    \includegraphics[width=\textwidth]{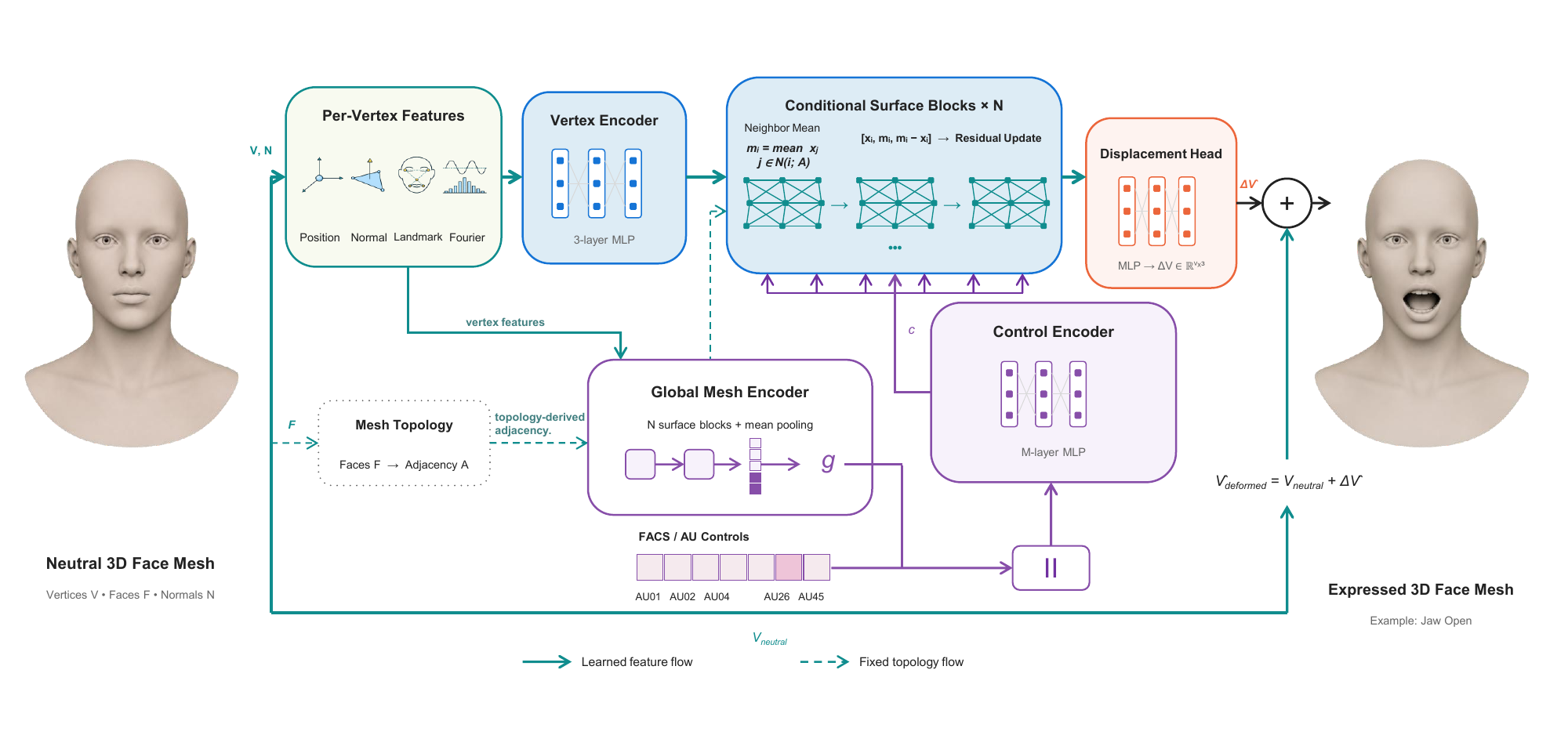}
    \caption{
    \textbf{TopoRig architecture.}
    Per-vertex geometric, landmark, and Fourier features are processed using
    the input mesh connectivity, while a global mesh representation and AU
    control jointly condition the deformation network. The model predicts a
    3D displacement for each input vertex, preserving the original topology.
    }
\label{fig:toporig_architecture}
\end{figure*}

Automatic facial rigging has traditionally relied on registering a
canonical template and transferring blendshape or skinning deformations
to a target mesh~\cite{carrigan2020expression,dutreve2010easy,onizuka2019landmark}, often requiring reliable correspondence and
per-character processing. Neural approaches instead learn deformation
mappings across identities and mesh structures.

Neural Face Rigging (NFR) introduced a deformation autoencoder trained using interpretable controls from a linear 3D morphable model and detailed deformations from 4D facial captures \cite{qin2023nfr}. This combination allows NFR to retain user-controllable expression parameters while reproducing fine-grained details from real facial performances.

Neural Face Skinning later combined a global expression representation with localized facial deformation \cite{cha2025nfs}. The method predicts per-vertex skinning weights that localize the influence of a global expression representation to different facial regions. These weights are learned using facial segmentation labels, while FACS-based blendshapes provide interpretable expression supervision.

AUBlendNet instead predicts 32 personalized AU blendshape bases from a FLAME-based neutral identity mesh \cite{li2025aublendnet,li2017flame}. Its accompanying AUBlendSet dataset contains artist-created AU blendshapes for 500 identities. Because all identities share the fixed FLAME topology of 5,023 vertices, AUBlendNet does not directly address facial meshes with arbitrary topology.

RigAnyFace directly predicts FACS-conditioned displacements on the vertices of an input neutral facial mesh using a deformation network built on DiffusionNet \cite{ma2025riganyface,sharp2022diffusionnet}. The DiffusionNet blocks are modified to incorporate FACS conditioning, while a global encoder allows information to be shared across disconnected components such as the face and eyeballs. To reduce its dependence on expensive artist-rigged data, RigAnyFace combines 3D supervision from professionally rigged meshes with image-based supervision generated for unrigged meshes. Since its predicted displacements are applied directly to the original vertices, the input topology is preserved.

Most recently, OmniFaceRig introduced a template-based pipeline that converts a static surface-only character into an inner-mouth-aware facial rig \cite{wang2026omnifacerig}. The resulting rig contains up to 155 FACS blendshapes together with procedurally fitted teeth, gums, and tongue. Its selected facial template is rigidly and non-rigidly registered to the input before the blendshapes and inner-mouth components are transferred. Unlike RigAnyFace, which predicts displacements directly on the input vertices and therefore preserves their connectivity, OmniFaceRig constructs the rigged facial region through template fitting and fusion. Consequently, although it accepts diverse input topologies, it does not preserve the original facial connectivity.

TopoRig is most closely related to RigAnyFace in that both predict
FACS-conditioned deformations directly on the vertices of the input mesh
and therefore preserve its original connectivity. Our focus, however, is
on constructing and combining complementary supervision from a single
canonical AU rig. We derive accurate but template-biased common-topology
targets, topology-diverse transferred targets on identity-specific meshes,
and targeted image-based supervision for controls where geometric transfer
is unreliable. TopoRig then learns a single amortized deformation model
from these heterogeneous sources, avoiding the per-identity fitting and
transfer procedure at inference time.

%% file: sections/3_method.tex
\section{Method}
\label{sec:method}

\subsection{Overview}
\label{sec:overview}

TopoRig is a topology-preserving facial rigging framework that predicts
expression-dependent deformations for a neutral mesh. Given neutral
vertex positions $\mathbf{V}$, vertex normals $\mathbf{N}$,
vertex landmark features $\mathbf{L}$, mesh connectivity
$\mathcal{T}$, and an action-unit control vector $\mathbf{c}_{\mathrm{AU}}$, the
network predicts a three-dimensional displacement, $\widehat{\Delta \mathbf{V}}$, for every input
vertex:
\begin{equation}
f_{\theta}
\left(
\mathbf{V},\mathbf{N},\mathbf{L},\mathcal{T},\mathbf{c}_{\mathrm{AU}}
\right)
=
\widehat{\Delta \mathbf{V}}.
\label{eq:toporig}
\end{equation}
The expressed mesh is obtained by adding the predicted displacements
to the neutral vertices:
\begin{equation}
\widehat{\mathbf{V}}_{\mathrm{expr}}
=
\mathbf{V}+\widehat{\Delta\mathbf{V}}.
\label{eq:expressed_mesh}
\end{equation}
The network combines local surface geometry, facial landmark features,
global facial geometry, and the requested expression. Since the
predicted displacements are applied directly to the original
vertices, TopoRig preserves the vertex count, faces, and connectivity
of the input mesh.

Here, \emph{topology-agnostic} means that TopoRig does not require a fixed
canonical topology, vertex ordering, or correspondence across input meshes;
the connectivity of each individual mesh is still used to define its local
neighborhoods. The overall architecture is illustrated in
Fig.~\ref{fig:toporig_architecture}.

\subsection{Per-Vertex Input Features}
\label{sec:inputs}

Let the neutral mesh contain $n_v$ vertices and $n_f$ triangular faces, where both values may vary between meshes. It is represented by vertex positions $\mathbf{V}\in\mathbb{R}^{n_v\times3}$, unit vertex normals $\mathbf{N}\in\mathbb{R}^{n_v\times3}$, and face indices $\mathcal{T}\in\mathbb{N}^{n_f\times3}$. 

TopoRig can optionally augment each vertex with facial landmark features that provide semantic information about nearby facial regions. A landmark detector provides a set of semantically consistent 3D facial locations with fixed indices. For each mesh vertex $i$, we select its $K_{\mathrm{L}}$ nearest landmarks. For each selected landmark, we encode five values: three coordinates for the 3D offset from the vertex to the landmark, one value for their Euclidean distance, and one normalized landmark index.  Concatenating these values over the $K_{\mathrm{L}}$ landmarks gives the feature $\mathbf{l}_i\in\mathbb{R}^{5K_{\mathrm{L}}}$. During training, we improve robustness to incomplete landmark
information using two forms of landmark dropout. Modality dropout
removes all landmark features, while region dropout removes landmarks
from a randomly selected facial region before the nearest-landmark
features are constructed. In both cases, the per-vertex input
dimensionality remains fixed.

The relative offsets, distances, and landmark indices provide semantic localization cues without requiring vertex correspondence, helping the network identify facial regions across meshes with different vertex orderings and connectivity.

We normalize the vertex coordinates of each mesh by subtracting the
mesh centroid and scaling by the maximum vertex distance from the
centroid, yielding $\widetilde{\mathbf{v}}_i$. To provide additional
spatial information, we apply a Fourier positional encoding with two frequency bands to these normalized coordinates
\cite{tancik2020fourier}. Let $\gamma(\widetilde{\mathbf{v}}_i)$ denote
this encoding. The complete per-vertex feature is then

\begin{equation}
\mathbf{z}_i
=
\left[
\widetilde{\mathbf{v}}_i,\,
\mathbf{n}_i,\,
\mathbf{l}_i,\,
\gamma(\widetilde{\mathbf{v}}_i)
\right]
\in\mathbb{R}^{58},
\label{eq:vertex_features}
\end{equation}

where $\mathbf{n}_i$ is the unit surface normal and
$\mathbf{l}_i$ is the landmark-relative feature defined above.

\subsection{Conditional Deformation Network}

TopoRig represents facial expressions using a 53-dimensional control
vector, denoted by $\mathbf{c}_{\mathrm{AU}}$, whose entries correspond to
the expression controls provided by ICT FaceKit. These controls follow the AU-based naming used by ICT FaceKit throughout this work. In all training and evaluation experiments, each target
expression activates a single control and is therefore represented by a
one-hot vector.

A vertex encoder $E_{\mathrm{v}}$, implemented as an MLP, maps each
per-vertex input feature $\mathbf{z}_i$ to an initial feature for the
deformation network,

\begin{equation}
\mathbf{x}_i^{(0)}
=
E_{\mathrm{v}}(\mathbf{z}_i).
\label{eq:vertex_encoder}
\end{equation}

In parallel, inspired by the pooled global representation used in
RigAnyFace~\cite{ma2025riganyface}, a separate global encoder
summarizes the geometry of the complete neutral mesh into a
mesh-level representation. While the
deformation branch operates primarily on local vertex neighborhoods,
this global representation provides identity-level shape context that
is shared across the entire mesh. Its input MLP
$E_{\mathrm{g,in}}$ maps the same per-vertex feature to

\begin{equation}
\mathbf{h}_i^{(0)} = E_{\mathrm{g,in}}(\mathbf{z}_i).
\end{equation}

The triangular faces induce a bidirectional vertex adjacency
$\mathbf{A}\in\{0,1\}^{n_v\times n_v}$, with self-edges included for
all vertices.

At layer $l$ of the global encoder, each vertex computes the mean
feature over itself and its adjacent vertices,
\begin{equation}
\mathbf{m}_{i,\mathrm{g}}^{(l)}
=
\frac{\sum_{j=1}^{n_v} A_{ij}\mathbf{h}_j^{(l)}}
{\sum_{j=1}^{n_v} A_{ij}}.
\end{equation}

The current feature, neighbourhood mean, and their difference are
combined by an MLP $E_{\mathrm{g}}^{(l)}$ and added residually to the
current representation,
\begin{align}
\mathbf{u}_i^{(l)}
&=
E_{\mathrm{g}}^{(l)}
\left(
[\mathbf{h}_i^{(l)},
 \mathbf{m}_{i,\mathrm{g}}^{(l)},
 \mathbf{m}_{i,\mathrm{g}}^{(l)}-\mathbf{h}_i^{(l)}]
\right),\\
\mathbf{h}_i^{(l+1)}
&=
\operatorname{LayerNorm}
\left(
\mathbf{h}_i^{(l)}+\mathbf{u}_i^{(l)}
\right).
\end{align}

After $L_{\mathrm{g}}$ such layers, an output MLP
$E_{\mathrm{g,out}}$ is applied to each vertex feature and the
resulting features are averaged to obtain a global representation
of the neutral identity,
\begin{equation}
\mathbf{g}
=
\frac{1}{n_v}
\sum_{i=1}^{n_v}
E_{\mathrm{g,out}}
\left(
\mathbf{h}_i^{(L_{\mathrm{g}})}
\right).
\end{equation}

The global representation $\mathbf{g}$ is concatenated with the AU
control vector $\mathbf{c}_{\mathrm{AU}}$ and passed through a control MLP
$E_{\mathrm{ctrl}}$ to produce the expression condition
\begin{equation}
\mathbf{c}
=
E_{\mathrm{ctrl}}([\mathbf{c}_{\mathrm{AU}},\mathbf{g}]).
\end{equation}

The condition $\mathbf{c}$ is then broadcast to every vertex in the
deformation network. In this way, each local deformation is
conditioned not only on the requested expression but also on the
global geometry of the input identity.

The main deformation network then applies $L_{\mathrm{d}}$
conditional aggregation layers, starting from the vertex features
$\mathbf{x}_i^{(0)}$. Using the same adjacency matrix, each layer
first computes

\begin{equation}
\mathbf{m}_i^{(l)}
=
\frac{
\sum_{j=1}^{n_v}
A_{ij}\mathbf{x}_j^{(l)}
}{
\sum_{j=1}^{n_v} A_{ij}
}.
\label{eq:neighbour_mean}
\end{equation}

The current vertex feature, its neighbourhood mean, and their
difference are then combined by an MLP
$E_{\mathrm{surf}}^{(l)}$ to form a local feature,

\begin{equation}
\mathbf{s}_i^{(l)}
=
E_{\mathrm{surf}}^{(l)}
\left(
[
\mathbf{x}_i^{(l)},
\mathbf{m}_i^{(l)},
\mathbf{m}_i^{(l)}-\mathbf{x}_i^{(l)}
]
\right).
\label{eq:surface_feature}
\end{equation}

The expression condition $\mathbf{c}$ is concatenated with this local
feature and processed by a second MLP
$E_{\mathrm{cond}}^{(l)}$. The resulting feature is added residually
to the current vertex representation,

\begin{equation}
\mathbf{x}_i^{(l+1)}
=
\operatorname{LayerNorm}
\left(
\mathbf{x}_i^{(l)}
+
E_{\mathrm{cond}}^{(l)}
\left(
[
\mathbf{s}_i^{(l)},\mathbf{c}
]
\right)
\right).
\label{eq:surface_block}
\end{equation}

This operation propagates information according to the input mesh
connectivity while conditioning each vertex on the requested
expression. After $L_{\mathrm{d}}$ conditional layers, the final vertex
feature is concatenated with the expression condition and passed
through a displacement MLP $E_{\mathrm{disp}}$,

\begin{equation}
\Delta\widehat{\mathbf{v}}_i
=
E_{\mathrm{disp}}
\left(
[
\mathbf{x}_i^{(L_{\mathrm{d}})},\mathbf{c}
]
\right).
\label{eq:displacement_head}
\end{equation}

The displacement head outputs a three-dimensional displacement for
each input vertex. Its final layer is initialized to zero, so the
untrained network initially predicts zero displacement and reproduces
the neutral input mesh.

\begin{figure}[t]
    \centering
    \includegraphics[width=\columnwidth]
    {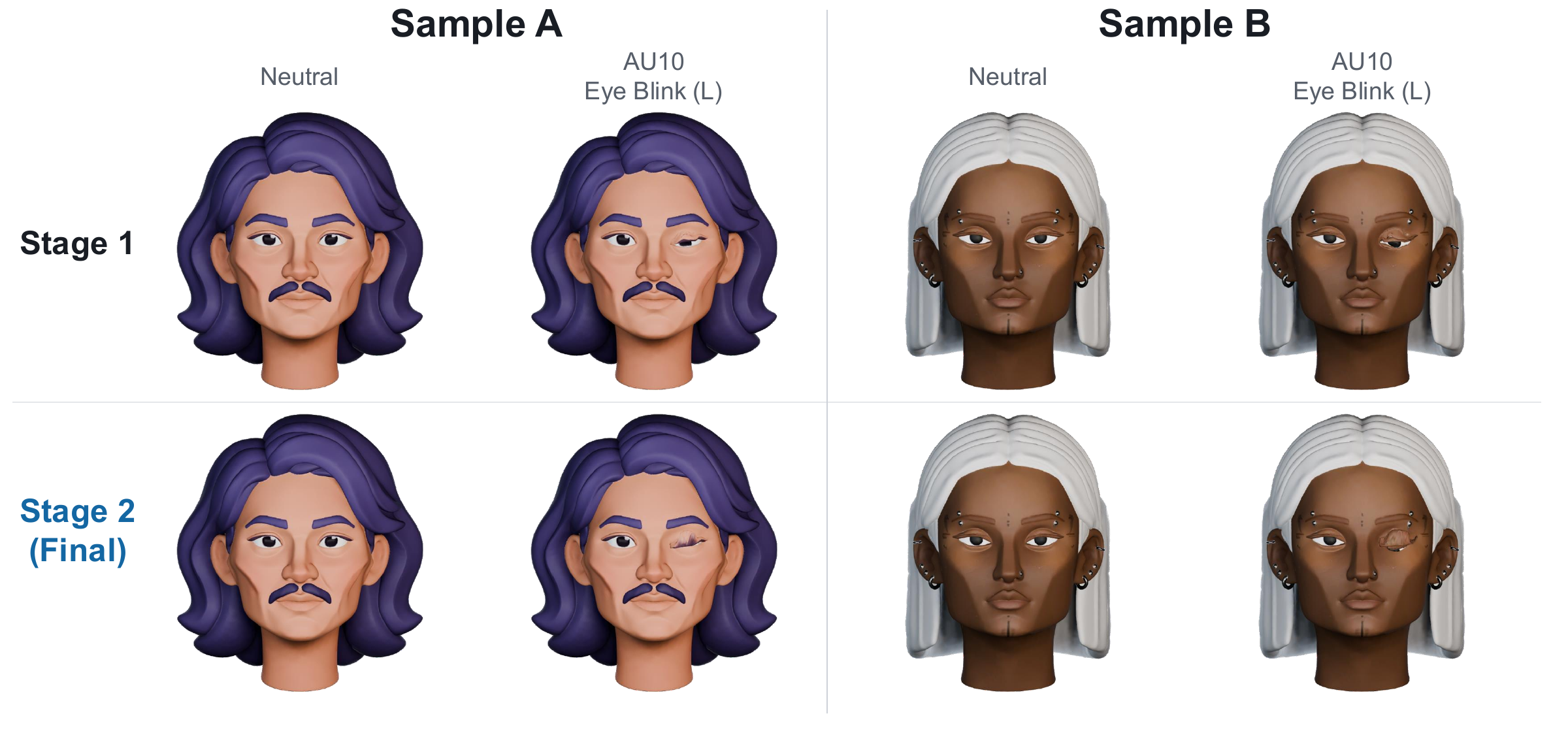}
    \caption{
    \textbf{Effect of Stage~2 image-guided refinement.}
    For left eye blink, Stage~1 inherits weak closure from unreliable
    transferred targets, while Stage~2 image supervision produces stronger
    and more complete eyelid closure on both validation identities.
    }
    \label{fig:toporig_stage_comparison}
\end{figure}

\subsection{Training Objectives}
\label{sec:training_objectives}

TopoRig is trained in two stages. Stage~1 learns the
FACS-conditioned deformation model from mesh-supervised samples.
Stage~2 initializes from Stage~1 and jointly uses mesh- and
image-supervised samples. For mesh-supervised samples, the primary vertex loss
$\mathcal{L}_{\mathrm{vtx}}$ supervises the predicted displacement
field against the target expression deformation. We additionally use
a landmark loss $\mathcal{L}_{\mathrm{lmk}}$ and an active-landmark
loss $\mathcal{L}_{\mathrm{active}}$ to increase supervision around
semantically important facial regions and landmarks associated with
the selected AU, respectively. The mesh objective is

\begin{equation}
\mathcal{L}_{\mathrm{mesh}}
=
\mathcal{L}_{\mathrm{vtx}}
+
\lambda_{\mathrm{lmk}}\mathcal{L}_{\mathrm{lmk}}
+
\lambda_{\mathrm{active}}^{\mathrm{eff}}
\mathcal{L}_{\mathrm{active}}
+
\mathcal{L}_{\mathrm{geom}}.
\end{equation}

The geometric regularizer $\mathcal{L}_{\mathrm{geom}}$ combines
displacement smoothness, relative edge-length preservation,
surface-normal consistency, and displacement-magnitude penalties:

\begin{align}
\mathcal{L}_{\mathrm{geom}}
={}&
\lambda_{\mathrm{smooth}}\mathcal{L}_{\mathrm{smooth}}
+
\lambda_{\mathrm{edge}}\mathcal{L}_{\mathrm{edge}}
\nonumber\\
&+
\lambda_{\mathrm{normal}}\mathcal{L}_{\mathrm{normal}}
+
\lambda_{\mathrm{mag}}\mathcal{L}_{\mathrm{mag}}.
\label{eq:geometry_loss}
\end{align}

Landmark features provide semantic localization, while the landmark
losses directly supervise deformation near expression-relevant facial
regions.

For image-supervised samples, target optical flow between the
neutral and expressed images is estimated using WAFT~\cite{wang2026waft}.
We apply image-based supervision only to a subset of eight
eye-region controls for which the transferred 3D blendshapes are
noisy or insufficiently reliable, particularly for eyelid closure
and opening.

In addition to optical flow, we use an eyelid objective
$\mathcal{L}_{\mathrm{eyelid}}$ that supervises normalized
projected eyelid motion and eyelid-gap reduction, together with a
localized depth-anchor term. We further regularize the predicted
deformation with an image-branch geometric regularizer
$\mathcal{L}_{\mathrm{geom}}^{\mathrm{img}}$ and an outside-region
anchor loss $\mathcal{L}_{\mathrm{anchor}}$ that suppresses motion
away from the supervised facial region. The image objective is
\begin{equation}
\mathcal{L}_{\mathrm{image}}
=
\lambda_{\mathrm{flow}}\mathcal{L}_{\mathrm{flow}}
+
\mathcal{L}_{\mathrm{eyelid}}
+
\mathcal{L}_{\mathrm{geom}}^{\mathrm{img}}
+
\lambda_{\mathrm{anchor}}\mathcal{L}_{\mathrm{anchor}}.
\end{equation}
During Stage 2, mesh- and image-supervised samples are trained jointly.
The loss applied to each sample depends on its available supervision:
\begin{equation}
    \mathcal{L}
    =
    \,\mathcal{L}_{\mathrm{mesh}}
    +
    \,\mathcal{L}_{\mathrm{image}}.
\end{equation}
Detailed definitions of the individual mesh and image loss terms are
provided in Sec.~\ref{sec:supp_losses} of the supplementary material. At inference time, the network directly predicts a displacement field
for a requested AU control on the neutral mesh. Since only the vertex
positions are modified, the original vertex count, faces, and
connectivity are preserved.

%% file: sections/4_experiments.tex
\section{Experiments}

\subsection{Datasets and Metrics}
\label{sec:datasets_metrics}

\noindent\textbf{Datasets.}
We use 3,496 facial identities generated using Nano
Banana~\cite{google2025nanobanana} and reconstruct their neutral meshes using
Pixal3D~\cite{li2026pixal3d}. The identities are divided into 2,796 training,
350 validation, and 350 test identities.

We obtain 3D supervision by fitting ICT FaceKit~\cite{li2020learning} to each reconstructed identity. This produces both a common-topology ICT mesh and an expression-transferred mesh that preserves the original Pixal3D topology. The original-topology meshes are simplified to approximately 100,000
faces (denoted \textbf{Pixal3D-100k}), and MediaPipe landmarks~\cite{kartynnik2019facemesh}
are transferred to the simplified geometry. Further construction details are provided in \cref{app:blendshape_transfer}. The full ICT FaceKit control vocabulary contains 53 expression controls.
For mesh-supervised experiments, we use the 45 non-gaze controls available
in our fitted ICT stream. Four of these have unreliable transferred targets
and are therefore omitted from the Pixal3D stream, leaving 41 controls.
Image supervision is applied to eight eye-region controls and uses 17,504
expression pairs from 2,779 identities.

The Pixal3D expression meshes are automatically constructed by transferring
ICT expression deformations to the original Pixal3D topology. We therefore
treat them as \emph{transferred reference targets}, rather than independent
ground-truth rigs. Evaluation on this set measures how well a method preserves
the canonical expression space while generalizing to previously unseen
identities and mesh topologies. Importantly, lower error does not necessarily
imply higher perceptual rig quality, since the transferred references may
contain local correspondence artifacts. We therefore use this evaluation as
a measure of transfer fidelity and topology generalization, complemented by
qualitative results on diverse character meshes.

The held-out evaluation set contains 350 identities, yielding
15,750 ICT and 14,350 Pixal3D mesh--expression samples.
All methods use the same identities, controls, preprocessing,
scale normalization, and evaluation masks.

\textbf{Evaluation Metrics.}
Predictions are evaluated as displacement fields rather than absolute
vertex positions. Each neutral head is standardized to a height of
240\,mm, and evaluation is restricted to vertices whose reference
displacement exceeds 0.01\,mm. We report coordinate-wise MAE and the
95th percentile (Q95) of per-vertex Euclidean displacement error,
averaged equally across identity--control pairs. MAE captures average
deformation error, while Q95 emphasizes larger localized deviations.

\subsection{Implementation Details}

We use $K_L=8$ nearest landmarks, two Fourier frequency bands, and two
aggregation layers in both the global encoder and deformation network.
Training uses AdamW with a global batch size of 16. Stage~1 is trained for
20 epochs with mesh supervision, while Stage~2 is initialized from Stage~1
and refined for 10 epochs using joint mesh and image supervision.
Image supervision is applied to eight eye-region controls, and Stage~2
additionally includes mesh-only refinement for selected higher-error controls.
Full optimization, loss-weight, augmentation, and refinement settings are
provided in the supplementary material.

\begin{figure*}[t]
    \centering
    \includegraphics[width=\textwidth]{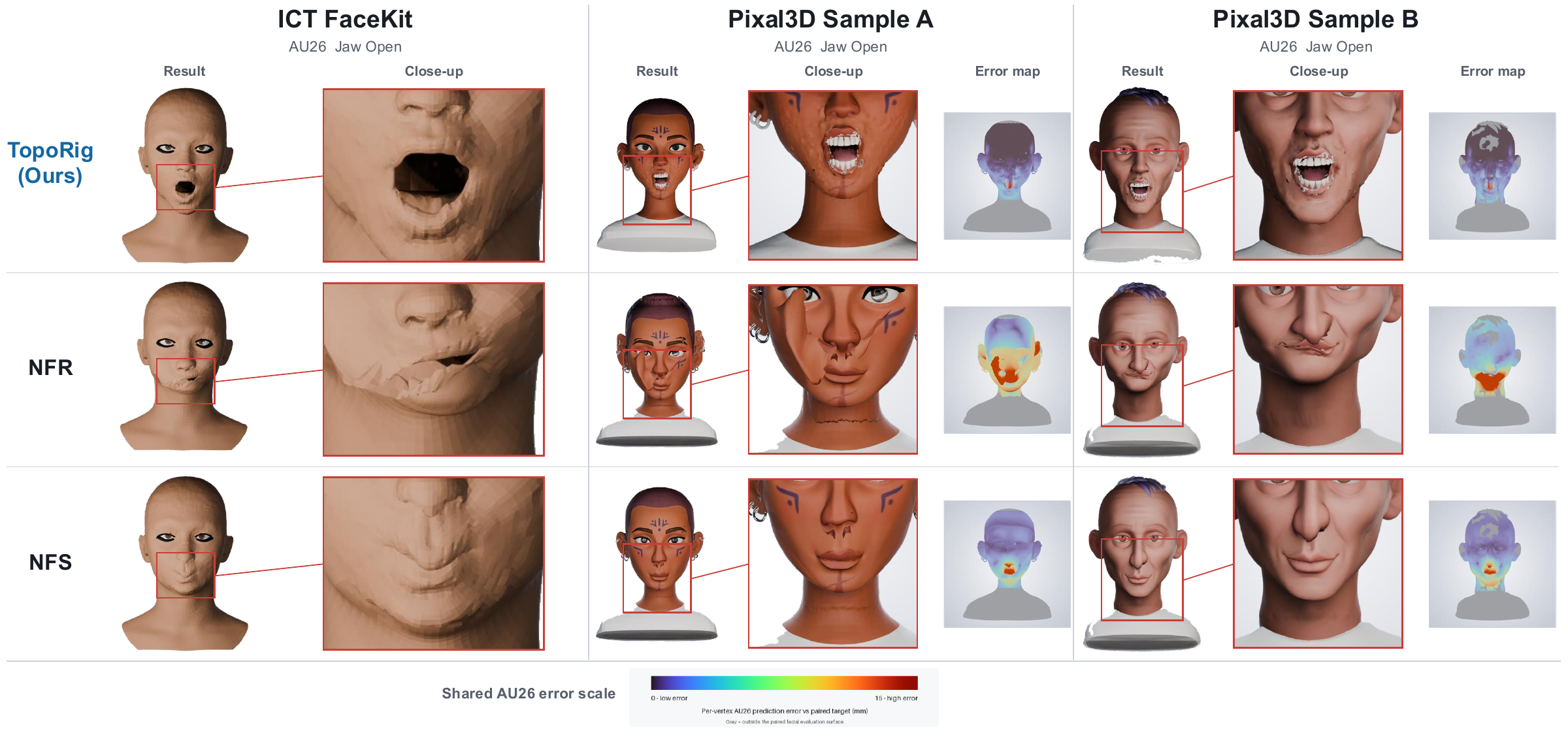}
    \caption{
    \textbf{Qualitative comparison with NFR and NFS for jaw opening.}
    Columns show the ICT FaceKit topology and two Pixal3D identities with
    custom topologies. For Pixal3D, error maps show per-vertex Euclidean
    distance to the transferred reference target on a shared $0$--$15$\,mm
    scale; gray denotes vertices outside the evaluated region. TopoRig
    produces more consistent deformation across common and custom topologies.
    }
    \label{fig:toporig_nfr_nfs_comparison}
\end{figure*}

\subsection{Comparison with baselines}
\label{sec:baseline_comparison}

\input{tables/1_quantitative_comp}

We compare TopoRig with Neural Face Rigging (NFR)~\cite{qin2023nfr} and Neural
Face Skinning (NFS)~\cite{cha2025nfs} using their publicly released pretrained
checkpoints. RigAnyFace~\cite{ma2025riganyface}, although closely related, is not included because
neither its implementation nor pretrained models are publicly available.
For each control, we provide the corresponding ICT expression together with
the neutral test mesh to each baseline. Since both baselines also predict a
neutral reconstruction, expression displacements are measured relative to
their predicted neutral output. All methods use the same test identities,
shared controls, preprocessing, evaluation masks, scale normalization, and
metrics described in~\cref{sec:datasets_metrics}.

As shown in Tab.~\ref{tab:baseline_comparison}, TopoRig substantially
outperforms both prior methods on common-topology ICT meshes and on
identity-specific Pixal3D meshes. Relative to the strongest baseline,
TopoRig reduces MAE by 72.9\% on ICT and 78.5\% on Pixal3D-100k,
with similarly large reductions in Q95. On Pixal3D-100k, the lower errors
indicate that TopoRig more faithfully reproduces the transferred reference
expressions across nonuniform, identity-specific topologies; they should
not be interpreted as an independent measure of perceptual rig quality.
Overall, TopoRig more accurately reproduces the reference expression
targets on previously unseen identities and mesh structures.

\begin{figure*}[t]
    \centering
    \includegraphics[width=\linewidth]
    {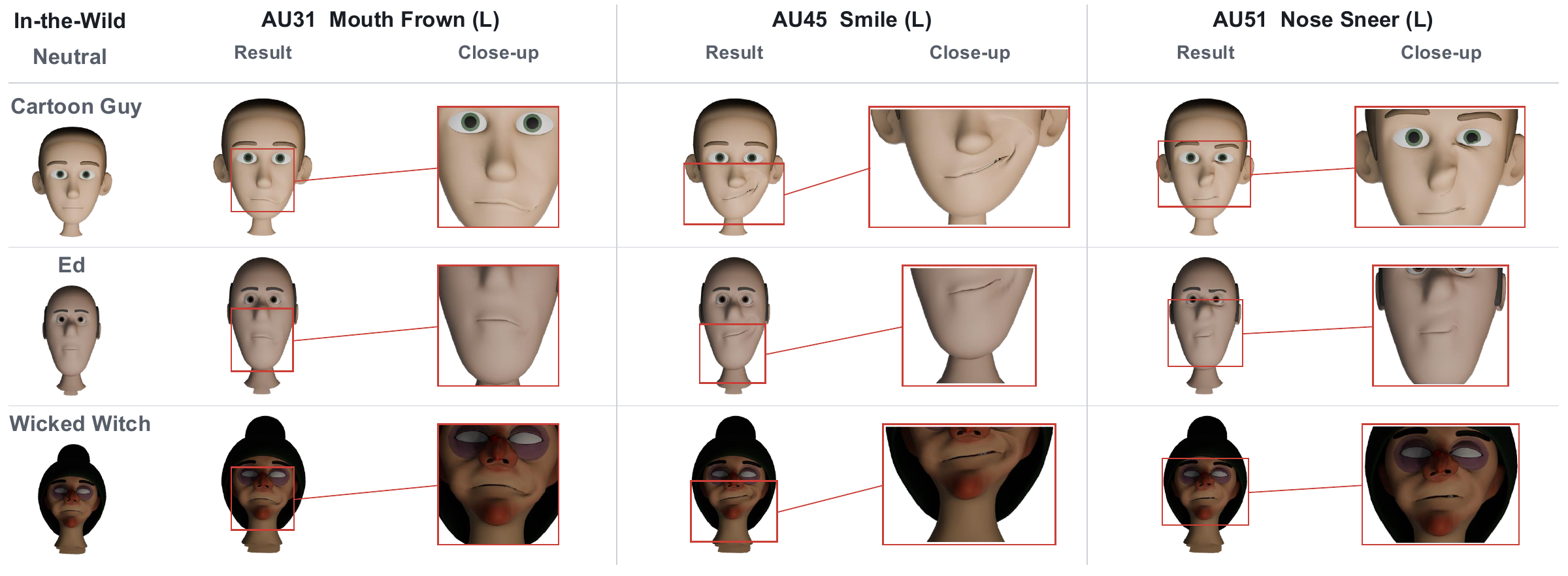}
    \caption{
    \textbf{Generalization to in-the-wild character meshes.}
    TopoRig produces localized AU-conditioned deformations on three
    stylized characters with geometry and appearance distinct from the
    training identities, while preserving their character-specific shape.
    The character meshes are publicly available CC0 assets from BlendSwap:
    \textit{Cartoon Guy V1.2} by mjedewaard, \textit{Ed} from
    \textit{Cartoon Character Pack 1} by VMComix, and
    \textit{Wicked Witch} by squibblejack. These meshes are used only for
    qualitative evaluation and are not part of the TopoRig training data.
    }
    \label{fig:in_the_wild}
\end{figure*}

Fig.~\ref{fig:toporig_nfr_nfs_comparison} provides a qualitative comparison
for jaw opening, showing more consistent deformation across the ICT template
and identity-specific Pixal3D topologies than NFR and NFS.
Fig.~\ref{fig:in_the_wild} further demonstrates generalization to stylized
in-the-wild characters whose facial geometry and appearance differ
substantially from the training identities. Additional qualitative results
across held-out validation identities and facial controls are provided in
Sec.~\ref{sec:supp_qualitative} of the supplementary material.

\subsection{Ablation and Training Analysis}
\label{sec:ablation}

\input{tables/2_ablation}

Tab.~\ref{tab:ablation} analyzes the input representation, the value of
complementary supervision and data diversity, and the effect of Stage~2
refinement.

Landmark-relative features are the dominant input cue.
Relative to Full Stage~1, removing them increases MAE by
54\% on ICT and 36\% on Pixal3D-100k, whereas removing
Fourier encoding causes only a small degradation. This indicates
that semantic localization is particularly important for
cross-topology generalization, with Fourier features providing
a complementary benefit.

Training on either the ICT or transferred Pixal3D supervision alone
produces clear domain-specific behavior, while joint training performs
well across both common and identity-specific topologies. Performance also improves consistently with
additional training identities, indicating that both complementary mesh
supervision and identity diversity support generalization.

Stage~2 changes aggregate MAE and Q95 only slightly because it targets a
small subset of controls with unreliable transferred 3D supervision.
Its benefit is more evident qualitatively: as shown in
Fig.~\ref{fig:toporig_stage_comparison}, image-guided refinement produces
stronger and more complete eyelid closure than mesh supervision alone.

\subsection{Discussion and Limitations}
\label{sec:discussion}

The results highlight a useful trade-off between the different supervision
sources used by TopoRig. Common-topology ICT targets provide consistent
expression supervision, but constrain the training geometry to a canonical
mesh structure. In contrast, transferred Pixal3D targets expose the model
to substantially more varied identity-specific geometry and connectivity,
at the cost of correspondence noise introduced by deformation transfer.
The single-source ablations in Tab.~\ref{tab:ablation} reflect this
trade-off: training on either source alone produces strong domain-specific
behavior, whereas joint training achieves substantially more balanced
performance across the two mesh domains. These results are consistent with the two supervision sources providing
complementary geometric biases, although the ablation does not fully
disentangle this effect from the additional supervision available under
joint training.

The evaluation should nevertheless be interpreted with several limitations.
Because the Pixal3D expression targets are themselves constructed by
automatic deformation transfer, they are not independently authored
ground-truth rigs. Quantitative performance on Pixal3D-100k therefore
measures agreement with the transferred canonical expression space and
generalization across unseen mesh structures, rather than perceptual rig
quality in isolation. The in-the-wild examples in
Fig.~\ref{fig:in_the_wild} provide additional qualitative evidence that
the learned deformation model can operate on character geometries outside
the generated training distribution, but these examples likewise lack
artist-authored target rigs and are therefore not used for quantitative
evaluation.

Our current expression supervision also focuses on individual AU controls:
all training and quantitative evaluation targets activate a single control.
Although the network accepts an AU control vector, compound expressions
have not been systematically supervised or evaluated. In addition,
image-guided refinement currently targets eight eye-region controls for
which transferred 3D supervision is particularly unreliable. Extending
this supervision to other difficult facial regions, evaluating combinations
of controls, and collecting artist-authored rigs or perceptual judgments
would provide stronger measures of animation quality beyond deformation
transfer fidelity.

%% file: tables/1_quantitative_comp.tex
\begin{table}[t]
    \centering
    \caption{
    \textbf{Quantitative comparison with prior facial-rigging methods.}
    Pixal3D-100k errors measure agreement with automatically transferred ICT
    reference expressions on held-out identities and topologies, rather than
    with independently authored ground-truth rigs.
    }
    \label{tab:baseline_comparison}

    \resizebox{\columnwidth}{!}{%
    \begin{tabular}{lcccc}
        \toprule
        \multirow{2}{*}{Method}
        & \multicolumn{2}{c}{ICT}
        & \multicolumn{2}{c}{Pixal3D-100k} \\
        \cmidrule(lr){2-3}
        \cmidrule(lr){4-5}
        & MAE (mm) $\downarrow$
        & Q95 (mm) $\downarrow$
        & MAE (mm) $\downarrow$
        & Q95 (mm) $\downarrow$ \\
        \midrule

        NFS \cite{cha2025nfs}
        & 1.476 & 4.358
        & 1.416 & 4.260 \\

        NFR \cite{qin2023nfr}
        & 1.928 & 5.480
        & 1.273 & 4.145 \\

        \textbf{TopoRig (Ours)}
        & \textbf{0.400} & \textbf{1.257}
        & \textbf{0.274} & \textbf{0.975} \\

        \bottomrule
    \end{tabular}%
    }
\end{table}

%% file: tables/2_ablation.tex
\begin{table}[t]
    \centering
    \caption{
    \textbf{Ablation and training analysis.}
    Effects of input features, supervision sources, training-set size, and
    Stage~2 refinement. Errors are computed on moving vertices after
    normalization to a 240\,mm head height relative to the held-out reference
    expression targets.
    }
    \label{tab:ablation}

    \resizebox{\columnwidth}{!}{%
    \begin{tabular}{lcccc}
        \toprule
        \multirow{2}{*}{Model}
        & \multicolumn{2}{c}{ICT}
        & \multicolumn{2}{c}{Pixal3D-100k} \\
        \cmidrule(lr){2-3}
        \cmidrule(lr){4-5}
        & MAE (mm) $\downarrow$
        & Q95 (mm) $\downarrow$
        & MAE (mm) $\downarrow$
        & Q95 (mm) $\downarrow$ \\
        \midrule

        No landmark input
        & 0.641 & 1.940
        & 0.373 & 1.456 \\

        No Fourier encoding
        & 0.446 & 1.438
        & 0.282 & 1.008 \\

        \midrule

        ICT only
        & \textbf{0.372} & \textbf{1.089}
        & 1.086 & 4.153 \\

        Pixal3D only
        & 1.165 & 3.974
        & 0.286 & 1.039 \\

        \midrule

        25\% data
        & 0.768 & 2.260
        & 0.351 & 1.255 \\

        50\% data
        & 0.512 & 1.737
        & 0.308 & 1.115 \\

        75\% data
        & 0.494 & 1.624
        & 0.291 & 1.044 \\

        \midrule

        Full Stage~1
        & 0.416 & 1.333
        & 0.275 & 0.983 \\

        \textbf{Full Stage~2 (TopoRig)}
        & 0.400 & 1.257
        & \textbf{0.274} & \textbf{0.975} \\

        \bottomrule
    \end{tabular}%
    }
\end{table}

%% file: sections/5_conclusion.tex
\section{Conclusion}

We presented TopoRig, a topology-agnostic facial rigging framework that
predicts AU-conditioned deformations directly on the vertices of an input
mesh while preserving its original topology. TopoRig learns from
complementary supervision sources comprising accurate common-topology
targets, topology-diverse transferred targets, and targeted image-based
cues for controls with unreliable 3D supervision.

Across held-out identities, TopoRig achieves substantially lower deformation
error than the evaluated neural baselines on both common-topology ICT meshes
and identity-specific Pixal3D meshes. Qualitative results demonstrate consistent localized deformations across
diverse identities and stylized in-the-wild character meshes. Our ablations show that landmark-relative
semantic features, complementary mesh supervision, and increased identity
diversity contribute to cross-topology generalization, while image-guided
refinement improves localized deformations such as eyelid closure.

The supervision sources play complementary roles: canonical-topology
targets provide consistent expression supervision, transferred meshes
introduce geometric and topological diversity, and targeted image cues
address regions where geometric transfer is unreliable. Overall, TopoRig
amortizes this heterogeneous supervision into a single topology-preserving
model, avoiding per-identity fitting and transfer while preserving the
connectivity and identity-specific structure of unseen character meshes.

%% file: sections/appendix.tex
\clearpage
\appendix
\setcounter{table}{0}
\setcounter{figure}{0}
\setcounter{page}{1}
\maketitlesupplementary

\renewcommand{\thesection}{\Alph{section}}

\renewcommand{\thetable}{S\arabic{table}}
\renewcommand{\thefigure}{S\arabic{figure}}

\input{sections/appendix/A_template_fitting}
\input{sections/appendix/B_training_loss}
\input{sections/appendix/C_eye_mouth_postprocess}
\input{sections/appendix/D_additional_qualitative}

%% file: sections/appendix/A_template_fitting.tex
\section{Template Fitting and Blendshape Transfer}
\label{app:blendshape_transfer}

To generate expression-capable training meshes while preserving the topology of each
target identity, we use the ICT FaceKit Light model~\cite{li2020learning}
as a canonical facial deformation template. ICT FaceKit provides a common facial
topology together with 53 expression blendshapes, allowing a consistent expression
space to be transferred across identities with different geometry and connectivity.

Given a neutral target mesh and the neutral ICT template, we first align the two
meshes using corresponding facial landmarks. The ICT template is then non-rigidly
deformed to match the target identity while retaining its original connectivity.
The resulting identity deformation is applied consistently to the ICT neutral mesh
and its expression shapes, preserving the expression displacement fields. Finally,
the fitted ICT blendshape displacements are spatially transferred onto the vertices
of the original target topology.

The overall procedure consists of four stages:
(i) landmark-based similarity alignment,
(ii) non-rigid ICT template fitting,
(iii) blendshape-preserving identity deformation, and
(iv) deformation-field transfer to the target topology.

\subsection{ICT FaceKit Expression Template}
\label{app:ict_template}

We use ICT FaceKit Light~\cite{li2020learning}, which provides a canonical
neutral facial mesh and 53 expression blendshapes registered to a common topology.
The model includes localized deformations of the brows, eyelids, cheeks, jaw, and
mouth, with asymmetric expressions represented independently for the left and right
sides of the face.

The ICT expression meshes share the same vertex ordering and connectivity as the
neutral template. We therefore represent each expression $k$ as a per-vertex
displacement field relative to the neutral mesh. Let $\mathbf{v}^{0}_i$ denote
vertex $i$ of the neutral ICT mesh and $\mathbf{v}^{k}_i$ denote the corresponding
vertex under expression $k$. The expression displacement is

\begin{equation}
    \boldsymbol{\delta}^{k}_i
    =
    \mathbf{v}^{k}_i - \mathbf{v}^{0}_i.
\end{equation}

These deformation fields form the canonical expression space that is subsequently
adapted to each target identity.

\subsection{Landmark-Based Similarity Alignment}
\label{app:blendshape_alignment}

Let

\begin{equation}
    \mathcal{M}_{\mathrm{ICT}}
    =
    (\mathbf{V}_{\mathrm{ICT}}, \mathbf{F}_{\mathrm{ICT}})
\end{equation}

denote the neutral ICT template and

\begin{equation}
    \mathcal{M}_{\mathrm{T}}
    =
    (\mathbf{V}_{\mathrm{T}}, \mathbf{F}_{\mathrm{T}})
\end{equation}

denote the neutral target mesh.

Before non-rigid fitting, the target mesh is placed into the coordinate frame of the
ICT template using corresponding facial landmarks. Given shared landmark positions
$\{\mathbf{p}^{\mathrm{ICT}}_l\}_{l=1}^{L}$ and
$\{\mathbf{p}^{\mathrm{T}}_l\}_{l=1}^{L}$, we estimate a similarity transformation
consisting of a scale $s$, rotation $\mathbf{R}$, and translation $\mathbf{t}$:

\begin{equation}
    \hat{\mathbf{p}}^{\mathrm{T}}_l
    =
    s\mathbf{p}^{\mathrm{T}}_l\mathbf{R}
    +
    \mathbf{t}.
\end{equation}

To improve robustness to inaccurate landmark correspondences, landmarks with large
residual errors are removed using percentile-based trimming and the transformation is
re-estimated using the remaining correspondences. The resulting transformation is
then applied to the complete target mesh.

\subsection{Non-Rigid ICT Template Fitting}
\label{app:nonrigid_fitting}

Following global alignment, the ICT template is non-rigidly deformed to match the
geometry of the target identity while retaining the original ICT mesh connectivity.
Let $\mathbf{v}^{0}_i$ denote the original position of ICT vertex $i$ and
$\mathbf{v}_i$ its current fitted position. The corresponding identity displacement is

\begin{equation}
    \mathbf{d}_i
    =
    \mathbf{v}_i-\mathbf{v}^{0}_i.
\end{equation}

The fitting procedure combines surface correspondence, semantic landmark constraints,
and topology-aware displacement smoothing.

\paragraph{Surface Correspondence.}

Each ICT vertex is associated with its nearest point on the aligned target surface.
The resulting correspondence term can be interpreted as

\begin{equation}
    \mathcal{L}_{\mathrm{surf}}
    =
    \sum_i
    \left\|
        \mathbf{v}_i -
        \operatorname{NN}_{\mathrm{T}}(\mathbf{v}_i)
    \right\|_2^2,
\end{equation}

where $\operatorname{NN}_{\mathrm{T}}(\mathbf{v}_i)$ denotes the nearest target
surface sample. Large-distance correspondences are removed using percentile-based
trimming. A lower-weight reverse correspondence from the target surface to the ICT
template is additionally used to encourage coverage of the target geometry.

\paragraph{Landmark Constraints.}

Nearest-surface correspondence alone does not enforce semantic alignment between
facial structures. We therefore constrain ICT landmark vertices to their
corresponding target landmark positions:

\begin{equation}
    \mathcal{L}_{\mathrm{lm}}
    =
    \sum_{l=1}^{L}
    w_l
    \left\|
        \mathbf{v}_{i_l} -
        \hat{\mathbf{p}}^{\mathrm{T}}_l
    \right\|_2^2.
\end{equation}

Landmark influence is propagated to nearby vertices using spatially decaying weights,
providing smooth semantic guidance around the eyes, mouth, nose, and brows.

\paragraph{Topology-Aware Smoothing.}

To maintain a coherent surface during fitting, neighboring ICT vertices are encouraged
to undergo similar identity displacements. For the one-ring neighborhood
$\mathcal{N}(i)$ of vertex $i$, the mean neighboring displacement is

\begin{equation}
    \bar{\mathbf{d}}_i
    =
    \frac{1}{|\mathcal{N}(i)|}
    \sum_{j\in\mathcal{N}(i)}
    \mathbf{d}_j.
\end{equation}

The corresponding regularization can be written as

\begin{equation}
    \mathcal{L}_{\mathrm{smooth}}
    =
    \sum_i
    \left\|
        \mathbf{d}_i -
        \bar{\mathbf{d}}_i
    \right\|_2^2.
\end{equation}

The fitting process can therefore be interpreted as balancing

\begin{equation}
    \mathcal{L}_{\mathrm{fit}}
    =
    \lambda_{\mathrm{surf}}\mathcal{L}_{\mathrm{surf}}
    +
    \lambda_{\mathrm{lm}}\mathcal{L}_{\mathrm{lm}}
    +
    \lambda_{\mathrm{smooth}}\mathcal{L}_{\mathrm{smooth}},
\end{equation}

although the implementation performs iterative weighted displacement updates rather
than explicitly optimizing a scalar objective through gradient descent.

\subsection{Geometric Refinement}
\label{app:fitting_refinement}

After the initial non-rigid registration, the fitted displacement field is further
refined to improve local surface correspondence. A higher-resolution sampling of the
target surface is used for local detail projection, while additional smoothing and
outlier removal suppress isolated artifacts.

Region-specific constraints are applied around sensitive structures such as the
eyelids and lips, where closely separated surfaces can otherwise result in incorrect
nearest-neighbor correspondences.

\subsection{Blendshape-Preserving Identity Deformation}
\label{app:blendshape_preservation}

Let $\mathbf{v}^{0}_i$ denote the neutral ICT vertex and $\mathbf{v}^{k}_i$ the same
vertex under expression $k$. The canonical expression displacement is

\begin{equation}
    \boldsymbol{\delta}^{k}_i
    =
    \mathbf{v}^{k}_i-\mathbf{v}^{0}_i.
\end{equation}

After fitting the neutral ICT template to the target identity, let the identity
deformation be

\begin{equation}
    \boldsymbol{\Delta}_i
    =
    \mathbf{v}^{\mathrm{fit}}_i-\mathbf{v}^{0}_i.
\end{equation}

The same identity deformation is applied to both the neutral template and every
expression shape:

\begin{equation}
    \hat{\mathbf{v}}^{0}_i
    =
    \mathbf{v}^{0}_i+\boldsymbol{\Delta}_i,
\end{equation}

\begin{equation}
    \hat{\mathbf{v}}^{k}_i
    =
    \mathbf{v}^{k}_i+\boldsymbol{\Delta}_i.
\end{equation}

Therefore,

\begin{align}
    \hat{\mathbf{v}}^{k}_i-\hat{\mathbf{v}}^{0}_i
    &=
    \left(\mathbf{v}^{k}_i+\boldsymbol{\Delta}_i\right)
    -
    \left(\mathbf{v}^{0}_i+\boldsymbol{\Delta}_i\right) \\
    &=
    \mathbf{v}^{k}_i-\mathbf{v}^{0}_i \\
    &=
    \boldsymbol{\delta}^{k}_i.
\end{align}

Thus, fitting changes the identity represented by the ICT template while retaining
its canonical expression deformation fields.

\subsection{Blendshape Transfer to the Target Topology}
\label{app:target_topology_transfer}

The fitted ICT mesh acts as an intermediate deformation template. To retain the
connectivity of the original target mesh, the ICT expression displacement fields are
transferred onto the target vertices.

For target vertex $\mathbf{q}_j$, a local set of neighboring fitted ICT vertices
$\mathcal{N}_{\mathrm{ICT}}(j)$ is identified. The transferred displacement for
expression $k$ is

\begin{equation}
    \boldsymbol{\delta}^{k}_{j,\mathrm{T}}
    =
    \sum_{i\in\mathcal{N}_{\mathrm{ICT}}(j)}
    w_{ji}\boldsymbol{\delta}^{k}_i.
\end{equation}

The unnormalized interpolation weights are defined using a Gaussian kernel,

\begin{equation}
    \tilde{w}_{ji}
    =
    \exp\left(
        -\frac{1}{2}
        \frac{d_{ji}^{2}}{\sigma^{2}}
    \right),
\end{equation}

where $d_{ji}$ denotes the Euclidean distance between target vertex $j$ and fitted
ICT vertex $i$. The normalized weights are

\begin{equation}
    w_{ji}
    =
    \frac{\tilde{w}_{ji}}
    {\sum_{m\in\mathcal{N}_{\mathrm{ICT}}(j)}
    \tilde{w}_{jm}}.
\end{equation}

Correspondences exceeding a maximum transfer distance are rejected. The resulting
target expression is then

\begin{equation}
    \mathbf{q}^{k}_j
    =
    \mathbf{q}^{0}_j
    +
    \boldsymbol{\delta}^{k}_{j,\mathrm{T}}.
\end{equation}

This allows the canonical ICT expression deformation space to be transferred to
target meshes with different vertex counts and connectivity.

\subsection{Semantic Expression Corrections}
\label{app:semantic_blendshape_corrections}

Spatial interpolation can produce ambiguous correspondences in regions containing
closely separated surfaces, particularly around the eyes and mouth. We therefore
apply expression-specific semantic constraints following deformation transfer.

For eye-related expressions, landmark-derived masks restrict transferred motion to
the relevant eyelid and surrounding facial regions. Additional constraints suppress
unintended deformation of nearby interior eye geometry, while blink expressions use
a localized eyelid closure correction.

Mouth-related expressions are similarly constrained using lip landmarks. In
particular, jaw-opening deformation is stabilized near the upper and lower lip
boundaries to reduce incorrect correspondence between nearby mouth surfaces.

\subsection{Resulting Training Meshes}
\label{app:fitting_output}

The final representation preserves the geometry and connectivity of the generated
target identity while inheriting the canonical ICT expression space:

\begin{equation}
\begin{aligned}
    &\text{Target identity}
    + \text{Target topology} \\
    &\quad + \text{transferred ICT expression blendshapes}.
\end{aligned}
\end{equation}

These expression-specific target meshes are subsequently used as mesh supervision
during training.

%% file: sections/appendix/B_training_loss.tex
\section{Training Loss Details}
\label{sec:supp_losses}

We provide the definitions of the mesh- and image-supervised loss
terms introduced in Sec.~3.4. Let $\Delta\hat{\mathbf v}_i$ and
$\Delta\mathbf v_i$ denote the predicted and target displacement of
vertex $i$, respectively, and let
$\hat{\mathbf v}_i=\mathbf v_i+\Delta\hat{\mathbf v}_i$.

\subsection{Mesh-Supervised Losses}

\paragraph{Vertex and landmark losses.}
We divide vertices into active and inactive sets according to the
target displacement,
\begin{equation}
\mathcal A=\{i:\|\Delta\mathbf v_i\|_2\geq\tau\},
\qquad
\mathcal I=\{i:\|\Delta\mathbf v_i\|_2<\tau\},
\end{equation}
with $\tau=10^{-3}$ in normalized coordinates. For a vertex set
$\mathcal S$, define
\begin{equation}
E(\mathcal S)=
\frac{1}{3|\mathcal S|}
\sum_{i\in\mathcal S}
\|\Delta\hat{\mathbf v}_i-\Delta\mathbf v_i\|_1.
\end{equation}
The primary vertex loss balances active and inactive regions and
normalizes by the target expression magnitude,
\begin{equation}
\mathcal L_{\mathrm{vtx}}
=
\frac{
\frac{1}{2}\left[E(\mathcal A)+E(\mathcal I)\right]
}{
\max\left(
\frac{1}{|\mathcal A|}
\sum_{i\in\mathcal A}\|\Delta\mathbf v_i\|_2,
10^{-3}
\right)
},
\end{equation}
where an empty region is omitted from the average.

Let $\mathcal K$ denote the MediaPipe landmark vertices and
$\mathcal K_{\mathrm{act}}=\mathcal K\cap\mathcal A$. The landmark
terms are
\begin{equation}
\mathcal L_{\mathrm{lmk}}=E(\mathcal K),
\qquad
\mathcal L_{\mathrm{active}}=E(\mathcal K_{\mathrm{act}}).
\end{equation}
The active-landmark coefficient is adaptively balanced as
\begin{equation}
\lambda_{\mathrm{active}}^{\mathrm{eff}}
=
\min\left(
20,\,
\frac{
5\,\mathcal L_{\mathrm{vtx}}
}{
\mathcal L_{\mathrm{active}}+\epsilon
}
\right),
\end{equation}
while $\lambda_{\mathrm{lmk}}=5$.

\paragraph{Geometric regularization.}
Let $\mathcal E$ and $\mathcal F$ denote the mesh edges and triangular
faces. We use
\begin{align}
\mathcal L_{\mathrm{smooth}}
&=
\frac{1}{3|\mathcal E|}
\sum_{(i,j)\in\mathcal E}
\|\Delta\hat{\mathbf v}_i-\Delta\hat{\mathbf v}_j\|_2^2,
\\
\mathcal L_{\mathrm{edge}}
&=
\frac{1}{|\mathcal E|}
\sum_{(i,j)\in\mathcal E}
\left(
\frac{\hat\ell_{ij}-\ell_{ij}}
{\max(\ell_{ij},10^{-4})}
\right)^2,
\\
\mathcal L_{\mathrm{normal}}
&=
\frac{1}{|\mathcal F|}
\sum_{f\in\mathcal F}
\left(1-\mathbf n_f^\top\hat{\mathbf n}_f\right),
\\
\mathcal L_{\mathrm{mag}}
&=
\frac{1}{n_v}
\sum_i
\left[
\max\left(
\|\Delta\hat{\mathbf v}_i\|_2-0.12,0
\right)
\right]^2,
\end{align}
where $\ell_{ij}$ and $\hat\ell_{ij}$ are the neutral and deformed
edge lengths, and $\mathbf n_f$ and $\hat{\mathbf n}_f$ are the
corresponding unit face normals. These terms form
$\mathcal L_{\mathrm{geom}}$ as defined in Eq.~(17); their weights
are reported in Sec.~4.2.

\subsection{Image-Supervised Losses}
\label{sec:supp_image_losses}

\paragraph{Optical-flow loss.}
WAFT provides a target flow $\mathbf F^{\mathrm{tgt}}$, while
differentiable projection of the neutral and predicted meshes gives
$\mathbf F^{\mathrm{pred}}$. For supervision mask $M$, we use the
masked Charbonnier loss
\begin{equation}
\mathcal L_{\mathrm{flow}}
=
\frac{
\sum_p M_p
\sum_{c\in\{x,y\}}
\left(
\sqrt{
(F^{\mathrm{pred}}_{p,c}-F^{\mathrm{tgt}}_{p,c})^2+\epsilon^2
}
-\epsilon
\right)
}{
2\sum_p M_p
},
\end{equation}
with $\epsilon=0.01$.

\paragraph{Eyelid objective.}
Let $\mathbf p_k^0,\mathbf p_k^1$ denote the neutral and
deformed projected mesh landmarks and
$\mathbf q_k^0,\mathbf q_k^1$ the corresponding target landmarks.
For image-derived targets, $\mathbf q_k^0$ and $\mathbf q_k^1$
are detected in the neutral and expressed images, respectively.
Each domain is normalized by its neutral eye width,
\begin{equation}
\hat{\mathbf m}_k
=
\frac{\mathbf p_k^1-\mathbf p_k^0}{w_{\mathrm{mesh}}},
\qquad
\mathbf m_k
=
\frac{\mathbf q_k^1-\mathbf q_k^0}{w_{\mathrm{img}}}.
\end{equation}
We subtract the mean motion of the two eye corners from both
quantities before comparison.

For paired upper--lower eyelid landmarks, we additionally measure
the reduction in eyelid gap. Let $g_k^0,g_k^1$ denote the neutral
and deformed projected mesh gaps and
$\tilde g_k^0,\tilde g_k^1$ the corresponding target gaps. We use
\begin{equation}
\hat r_k
=
\frac{g_k^0-g_k^1}{w_{\mathrm{mesh}}},
\qquad
r_k
=
\frac{\tilde g_k^0-\tilde g_k^1}{w_{\mathrm{img}}}.
\end{equation}

Let $\rho_\beta$ denote the Smooth-L1 loss with transition
parameter $\beta$. The projected-motion and gap terms are
\begin{equation}
\mathcal L_{\mathrm{motion}}
=
\rho_{0.05}(\hat{\mathbf m},\mathbf m),
\qquad
\mathcal L_{\mathrm{gap}}
=
\rho_{0.05}(\hat{\mathbf r},\mathbf r).
\end{equation}

For the two blink controls, we replace the detected expressed
landmarks with a synthetic full-closure target constructed from
the neutral projected mesh. Each paired upper eyelid landmark is
moved $0.8$ of the way toward its corresponding lower landmark,
while the lower landmark is moved $0.2$ of the way toward the
upper landmark.

\paragraph{Depth anchor.}
Because projected supervision alone does not constrain motion along
the depth direction, we penalize depth displacement within a
localized eyelid region. Let $\mathcal R_{\mathrm{depth}}$ denote
the corresponding mesh-hop region and let $z_i^0,z_i^1$ be the
neutral and deformed depth coordinates. We define
\begin{equation}
\mathcal L_{\mathrm{depth}}
=
\frac{1}{|\mathcal R_{\mathrm{depth}}|}
\sum_{i\in\mathcal R_{\mathrm{depth}}}
\rho_{0.01}
\left(
\frac{z_i^1-z_i^0}{w_{\mathrm{mesh}}},
0
\right).
\end{equation}

The complete eyelid objective is
\begin{equation}
\mathcal L_{\mathrm{eyelid}}
=
\lambda_{\mathrm{eyelid}}
\left(
\mathcal L_{\mathrm{motion}}
+
2\mathcal L_{\mathrm{gap}}
+
2\mathcal L_{\mathrm{depth}}
\right),
\end{equation}
where $\lambda_{\mathrm{eyelid}}=3$ for image-derived targets and
$\lambda_{\mathrm{eyelid}}=5$ for the two blink controls.

\paragraph{Image-space geometric regularization.}
We additionally penalize overall displacement magnitude,
\begin{equation}
\mathcal L_{\mathrm{disp}}
=
\frac{1}{3n_v}
\sum_i
\|\Delta\hat{\mathbf v}_i\|_2^2,
\end{equation}
and reuse the displacement-smoothness, relative edge-length, and
normal-consistency terms defined in Sec.~B.1:
\begin{equation}
\begin{aligned}
\mathcal L_{\mathrm{geom}}^{\mathrm{img}}
={}&
\lambda_{\mathrm{disp}}\mathcal L_{\mathrm{disp}}
+
\lambda_{\mathrm{smooth}}\mathcal L_{\mathrm{smooth}}
\\
&+
\lambda_{\mathrm{edge}}\mathcal L_{\mathrm{edge}}
+
\lambda_{\mathrm{normal}}\mathcal L_{\mathrm{normal}}.
\end{aligned}
\end{equation}

Finally, letting $\mathcal R$ denote the supervised facial region,
the outside-region anchor is
\begin{equation}
\mathcal L_{\mathrm{anchor}}
=
\frac{1}{|\bar{\mathcal R}|}
\sum_{i\notin\mathcal R}
\|\Delta\hat{\mathbf v}_i\|_2^2.
\end{equation}
The complete image objective is given in Eq.~(18) of the main
paper.

\subsection{Training and Implementation Settings}
\label{sec:supp_implementation}

\paragraph{Architecture and optimization.}
The per-vertex input has 58 channels: normalized position,
surface normal, 40 landmark-relative features from the
$K_L=8$ nearest MediaPipe landmarks, and a 12-dimensional
Fourier encoding with two frequency bands. Landmark indices are
normalized by $467$. Both the global encoder and deformation
network use two aggregation layers. We use LayerNorm and SiLU
activations with dropout $0.3$, and initialize the final
displacement layer to zero.

Training uses AdamW with zero weight decay, a global batch size of
16, and gradient clipping at $1.0$. Stage~1 is trained for
20 epochs with initial learning rate $5\times10^{-4}$.
Stage~2 initializes from the Stage~1 checkpoint and is trained for
10 epochs with initial learning rate $10^{-4}$. Both stages use
cosine learning-rate decay to $10^{-6}$. Landmark modality dropout
and landmark-region dropout are applied in both stages with
probabilities $0.15$ and $0.10$, respectively; the latter randomly
drops the left-eye, right-eye, or mouth landmarks.

\paragraph{Mesh-supervised losses.}
The active-motion threshold is $10^{-3}$ in normalized mesh
coordinates. The landmark loss weight is
$\lambda_{\mathrm{lmk}}=5$. The active-landmark term is adaptively
balanced relative to $\mathcal L_{\mathrm{vtx}}$ with a target
loss ratio of $5$ and a maximum effective weight of $20$.
The displacement-smoothness, relative edge-length, and
normal-consistency weights are $0.05$, $0.001$, and $0.0005$,
respectively. The displacement-magnitude term has weight $5$ and
penalizes displacement exceeding $0.12$ normalized units.

\paragraph{Image-supervised losses.}
Image supervision is applied to eight eye-region controls.
WAFT target flow is computed at $512\times512$, while predicted
mesh motion is rendered at $128\times128$ using nvdiffrast.
The flow loss has weight $0.1$.

For the image-branch geometric regularizer, the displacement,
displacement-smoothness, relative edge-length, and
normal-consistency weights are $0.05$, $1.5$, $0.05$, and $0.05$,
respectively. The outside-region anchor has weight $500$.
The supervised region extends eight mesh hops around the relevant
landmarks and twelve hops for cheek-squint controls. The eyelid
depth anchor uses an eight-hop region. Before image-space
supervision is evaluated, predicted displacements at coincident
vertices are synchronized using a tolerance of $10^{-6}$.

\begin{figure*}[!t]
    \centering
    \includegraphics[width=\linewidth]
    {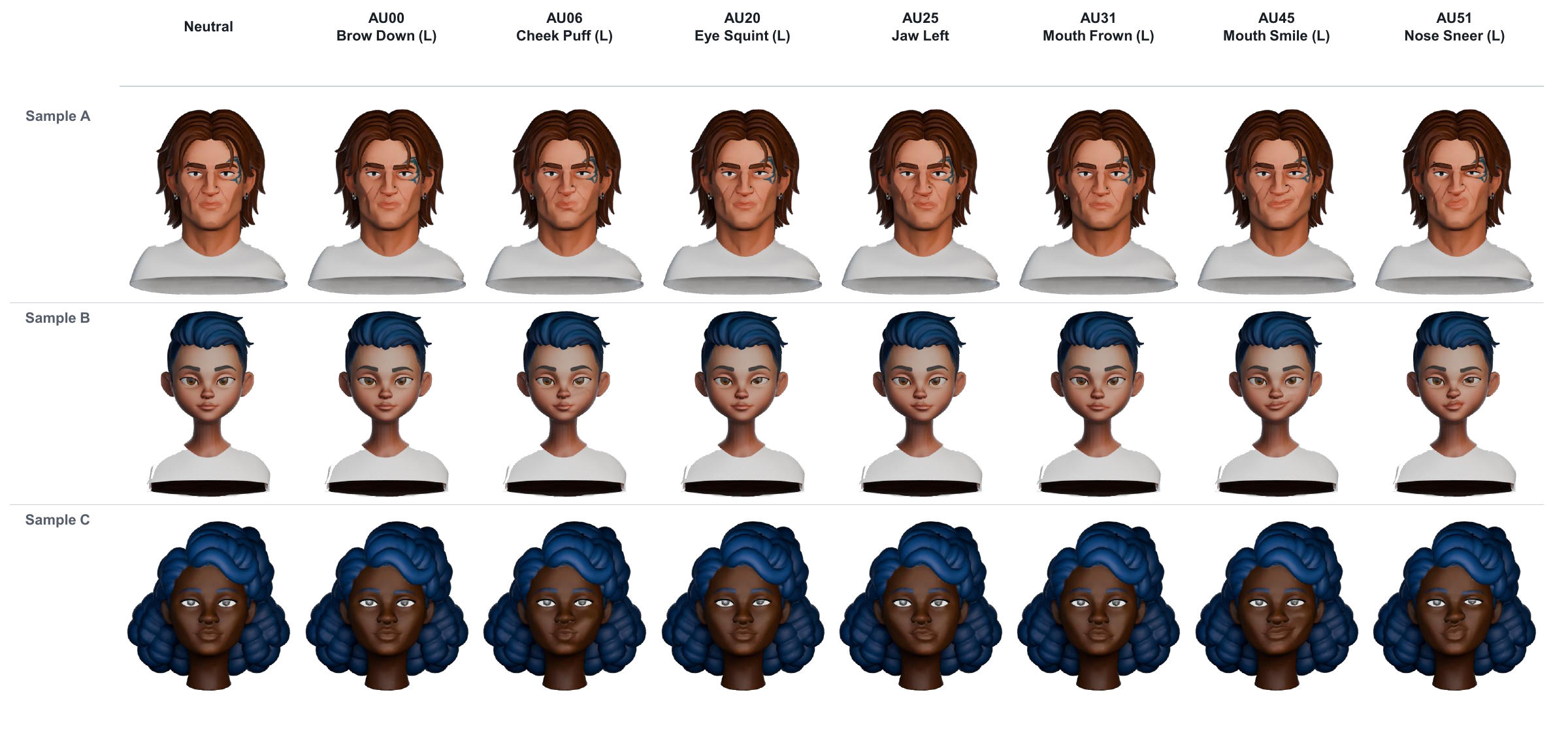}
    \caption{
        \textbf{Additional AU-conditioned expression results.}
        Rows show three held-out validation identities and columns show
        different unilateral and asymmetric facial controls. TopoRig produces
        localized deformations across diverse facial regions while preserving
        identity-specific geometry.
    }
    \label{fig:toporig_expression_results}
\end{figure*}

\paragraph{Stage~2 refinement.}
The mesh and image objectives have equal outer weights.
Stage~2 additionally includes mesh-only refinement streams from
both 3D supervision sources for AUs~24, 25, 26, 27, 28, 33, 34,
and 40, with the same unit outer loss weight. These streams provide
additional training exposure to controls with comparatively high
Stage~1 validation error and do not receive image supervision.

The checkpoint with the lowest mesh-only validation MAE is retained;
image losses are excluded from checkpoint selection.

%% file: sections/appendix/C_eye_mouth_postprocess.tex
\section{Eye and Inner-Mouth Post-processing}
\label{sec:supp_postprocessing}

TopoRig predicts per-vertex displacements on the input facial surface and is
therefore primarily designed for deformable structures such as the skin,
lips, and eyelids. Some facial components, however, are more naturally
described by rigid or articulated motion. In particular, eyeball rotation
should preserve the geometry of the eye, while the teeth should remain
approximately rigid during jaw motion. Furthermore, the generated facial
meshes used in our pipeline do not generally contain a complete inner-mouth
region. We therefore complement the learned facial deformation with
geometry-aware post-processing for eye-gaze and inner-mouth components. These
operations are separate from the learned TopoRig deformation model and do not
modify the connectivity of the original facial surface.

\paragraph{Eye-gaze motion.}
The 45 expression controls used to train TopoRig exclude the eight eye-gaze
controls AU12--AU19. We instead model these controls using rigid eyeball
rotation. We first localize the eye using facial and iris landmarks together
with their corresponding surface anchors. Connected components around the
iris are examined to distinguish eye geometry from the surrounding face and
to identify components belonging to the same physical eye. This procedure
supports assets in which the iris, cornea, sclera, or other visible eye
layers are represented by separate components. An approximately spherical
support is estimated from the selected geometry and used to determine the
eyeball center and radius.

When the source mesh contains only partial eye geometry, or when the visible
eye surface is fused with the surrounding facial mesh, we construct a complete
eyeball representation and transfer the source appearance to it. Additional
scleral backing is introduced when necessary to avoid exposing missing
geometry during rotation. The resulting representation allows gaze to be
modeled as a rigid transformation while retaining the appearance of the
original asset.

Let $\mathbf{m}\in\mathbb{R}^{2}$ denote the desired image-plane gaze motion,
and let $\mathbf{d}_{\mathrm{AU}}\in\mathbb{R}^{2}$ denote the unit direction
associated with the requested gaze control. We first extract the displacement
along the corresponding semantic gaze direction,
\begin{equation}
    s = \mathbf{m}^{\top}\mathbf{d}_{\mathrm{AU}} .
\end{equation}
Given an estimated eyeball radius $r$ and a maximum admissible rotation
$\theta_{\max}$, the displacement is bounded by
\begin{equation}
    \bar{s}
    =
    \operatorname{clip}
    \left(
        s,\,
        -r\sin\theta_{\max},\,
        r\sin\theta_{\max}
    \right),
\end{equation}
and converted to an angular displacement,
\begin{equation}
    \theta
    =
    \arcsin\left(\frac{\bar{s}}{r}\right).
\end{equation}
The corresponding normalized gaze intensity is
\begin{equation}
    \alpha
    =
    \operatorname{clip}
    \left(
        \frac{\theta}{\theta_{\max}},
        -1,1
    \right).
\end{equation}

For geometric application of the gaze motion, let $\mathbf{c}\in\mathbb{R}^3$
denote the estimated eyeball center and $\mathbf{p}\in\mathbb{R}^3$ the
neutral iris-center position. The initial viewing direction is
\begin{equation}
    \mathbf{u}
    =
    \frac{\mathbf{p}-\mathbf{c}}
         {\|\mathbf{p}-\mathbf{c}\|_2}.
\end{equation}
A calibrated image-plane gaze displacement is lifted to a three-dimensional
displacement $\Delta\mathbf{p}$. Since rigid rotation about the eyeball center
cannot produce radial translation, we retain only its component tangent to the
eyeball,
\begin{equation}
    \Delta\mathbf{p}_{\mathrm{tan}}
    =
    \Delta\mathbf{p}
    -
    \left(
        \Delta\mathbf{p}^{\top}\mathbf{u}
    \right)\mathbf{u}.
\end{equation}
The target viewing direction is subsequently defined as
\begin{equation}
    \mathbf{v}
    =
    \frac{
        r\mathbf{u}+\Delta\mathbf{p}_{\mathrm{tan}}
    }{
        \left\|
            r\mathbf{u}+\Delta\mathbf{p}_{\mathrm{tan}}
        \right\|_2
    }.
\end{equation}
We obtain the rigid rotation mapping $\mathbf{u}$ to $\mathbf{v}$ using
Rodrigues' formula. Defining
\begin{equation}
    \mathbf{k}=\mathbf{u}\times\mathbf{v},
    \qquad
    s_r=\|\mathbf{k}\|_2,
    \qquad
    c_r=\mathbf{u}^{\top}\mathbf{v},
\end{equation}
the rotation matrix is
\begin{equation}
    \mathbf{R}
    =
    \mathbf{I}
    +
    [\mathbf{k}]_{\times}
    +
    [\mathbf{k}]_{\times}^{2}
    \frac{1-c_r}{s_r^{2}},
\end{equation}
where $[\mathbf{k}]_{\times}$ denotes the skew-symmetric matrix associated
with $\mathbf{k}$. For negligible angular displacement, we use the identity
rotation. The selected eye components are transformed jointly using
$\mathbf{R}$, preserving their relative geometry. For AU12--AU19, the
network-predicted eye deformation is suppressed and gaze is produced by this
rigid-eye procedure, avoiding simultaneous learned and rigid motion.

\paragraph{Inner-mouth completion.}
The generated facial meshes primarily model the externally visible facial
surface and generally lack a complete inner-mouth region, including explicit
dental and interior oral geometry. This limitation is largely inconspicuous
for expressions in which the mouth remains closed, but becomes apparent for
large mouth-opening motions such as jaw opening (AU26), where the absence of
internal geometry produces an unrealistic empty cavity. We therefore insert
an auxiliary inner-mouth mesh containing the interior mouth surface together
with upper and lower dental geometry. The added component is positioned
relative to the surrounding facial landmarks so that it is aligned with the
mouth opening while remaining separate from the original facial surface.
Consequently, the connectivity of the input facial mesh is unchanged; the
inner mouth is introduced as an additional geometric component for the final
animated asset.

For source assets that already contain incomplete, duplicated, or otherwise
incompatible internal oral geometry, these components are removed or ignored
before inner-mouth completion. This prevents overlapping teeth or internal
surfaces from becoming visible when the jaw opens.

\paragraph{Dental motion.}
The inserted dental geometry is not independently deformed using the
per-vertex displacement field. Such deformation can cause individual tooth
vertices to move differently and thereby introduce non-rigid distortions.
Instead, the upper and lower dentition are treated as rigid components whose
motion is coupled to the facial deformation predicted by TopoRig. The upper
teeth remain fixed with respect to the upper face, whereas the lower teeth
follow the predicted motion of the lower jaw.

Let $\{\mathbf{x}_i\}_{i=1}^{N}$ denote neutral landmark positions selected
around the lower jaw, and let
\begin{equation}
    \hat{\mathbf{x}}_i
    =
    \mathbf{x}_i+\Delta\mathbf{x}_i
\end{equation}
denote their positions after applying the TopoRig-predicted displacement
field. We estimate the rigid transformation
$(\mathbf{R}_{j},\mathbf{t}_{j})$ that best aligns the neutral landmarks with
their predicted positions,
\begin{equation}
    (\mathbf{R}_{j}^{*},\mathbf{t}_{j}^{*})
    =
    \arg\min_{\mathbf{R}_{j}\in SO(3),\,\mathbf{t}_{j}}
    \sum_{i=1}^{N}
    \left\|
        \mathbf{R}_{j}\mathbf{x}_i
        +\mathbf{t}_{j}
        -\hat{\mathbf{x}}_i
    \right\|_2^2 .
\end{equation}
The resulting transformation is then applied uniformly to every lower-tooth
vertex $\mathbf{q}_j$,
\begin{equation}
    \hat{\mathbf{q}}_j
    =
    \mathbf{R}_{j}^{*}\mathbf{q}_j
    +
    \mathbf{t}_{j}^{*},
\end{equation}
while the upper dentition remains attached to the upper face. In this way,
the motion of the inserted inner-mouth geometry remains coupled to the
deformation predicted for the surrounding facial surface while preserving
the approximately rigid structure of the teeth.

%% file: sections/appendix/D_additional_qualitative.tex
\section{Additional Qualitative Results}
\label{sec:supp_qualitative}

Fig.~\ref{fig:toporig_expression_results} provides additional qualitative
results on held-out validation identities. We visualize several unilateral
and asymmetric facial controls spanning the brows, cheeks, eyes, jaw,
mouth, and nose. Across identities with substantially different facial
geometry and appearance, TopoRig produces localized AU-conditioned
deformations while preserving the characteristic shape of each input
identity.

%% file: main.bib
@article{cha2025nfs,
  author  = {Cha, Sihun and Yoon, Serin and Seo, Kwanggyoon and Noh, Junyong},
  title   = {Neural Face Skinning for Mesh-Agnostic
             Facial Expression Cloning},
  journal = {Computer Graphics Forum},
  volume  = {44},
  number  = {2},
  pages   = {e70009},
  year    = {2025},
  doi     = {10.1111/cgf.70009}
}

@book{ekman1978facs,
  author    = {Ekman, Paul and Friesen, Wallace V.},
  title     = {Facial Action Coding System: A Technique for the
               Measurement of Facial Movement},
  publisher = {Consulting Psychologists Press},
  address   = {Palo Alto, California},
  year      = {1978}
}

@misc{google2025nanobanana,
  author       = {Gerstenhaber, Michael},
  title        = {Building on the Bananas Momentum of Generative Media
                  Models on Google Cloud},
  howpublished = {Google Cloud Blog},
  year         = {2025},
  url          = {https://cloud.google.com/blog/products/ai-machine-learning/building-momentum-for-gen-media-including-nano-banana-}
}

@article{kartynnik2019facemesh,
  author  = {Kartynnik, Yury and Ablavatski, Artsiom and
             Grishchenko, Ivan and Grundmann, Matthias},
  title   = {Real-Time Facial Surface Geometry from Monocular Video
             on Mobile {GPUs}},
  journal = {arXiv preprint arXiv:1907.06724},
  year    = {2019},
  url     = {https://arxiv.org/abs/1907.06724}
}

@article{li2017flame,
  author  = {Li, Tianye and Bolkart, Timo and Black, Michael J. and
             Li, Hao and Romero, Javier},
  title   = {Learning a Model of Facial Shape and Expression
             from 4D Scans},
  journal = {ACM Transactions on Graphics},
  volume  = {36},
  number  = {6},
  year    = {2017},
  doi     = {10.1145/3130800.3130813}
}

@inproceedings{li2020learning,
  author    = {Li, Ruilong and Bladin, Karl and Zhao, Yajie and
               Chinara, Chinmay and Ingraham, Owen and Xiang, Pengda and
               Ren, Xinglei and Prasad, Pratusha and Kishore, Bipin and
               Xing, Jun and Li, Hao},
  title     = {Learning Formation of Physically-Based Face Attributes},
  booktitle = {Proceedings of the IEEE/CVF Conference on Computer Vision
               and Pattern Recognition (CVPR)},
  pages     = {3410--3419},
  year      = {2020},
  doi       = {10.1109/CVPR42600.2020.00347},
  url       = {https://arxiv.org/abs/2004.03458}
}

@inproceedings{li2025aublendnet,
  author    = {Li, Hao and Dai, Ju and Zhou, Feng and Ning, Kaida and
               Li, Lei and Pan, Junjun},
  title     = {{AU-Blendshape} for Fine-Grained Stylized
               3D Facial Expression Manipulation},
  booktitle = {Proceedings of the IEEE/CVF International Conference
               on Computer Vision (ICCV)},
  year      = {2025}
}

@inproceedings{li2026pixal3d,
  author    = {Li, Dong-Yang and Zhao, Wang and Chen, Yuxin and
               Hu, Wenbo and Guo, Meng-Hao and Zhang, Fang-Lue and
               Shan, Ying and Hu, Shi-Min},
  title     = {{Pixal3D}: Pixel-Aligned 3D Generation from Images},
  booktitle = {ACM SIGGRAPH 2026 Conference Papers},
  year      = {2026},
  doi       = {10.1145/3799902.3811175},
  url       = {https://arxiv.org/abs/2605.10922}
}

@inproceedings{ma2025riganyface,
  author    = {Ma, Wenchao and Kneubuehler, Dario and Chu, Maurice and
               Sachs, Ian and Jiang, Haomiao and Huang, Sharon X.},
  title     = {{RigAnyFace}: Scaling Neural Facial Mesh Auto-Rigging
               with Unlabeled Data},
  booktitle = {Advances in Neural Information Processing Systems},
  year      = {2025},
  url       = {https://wenchao-m.github.io/RigAnyFace.github.io/}
}

@inproceedings{qin2023nfr,
  author    = {Qin, Dafei and Saito, Jun and Aigerman, Noam and
               Groueix, Thibault and Komura, Taku},
  title     = {Neural Face Rigging for Animating and Retargeting
               Facial Meshes in the Wild},
  booktitle = {ACM SIGGRAPH 2023 Conference Proceedings},
  year      = {2023},
  doi       = {10.1145/3588432.3591556}
}

@article{sharp2022diffusionnet,
  author  = {Sharp, Nicholas and Attaiki, Souhaib and Crane, Keenan and
             Ovsjanikov, Maks},
  title   = {{DiffusionNet}: Discretization Agnostic Learning on Surfaces},
  journal = {ACM Transactions on Graphics},
  volume  = {41},
  number  = {3},
  year    = {2022},
  doi     = {10.1145/3507905}
}

@inproceedings{tancik2020fourier,
  author    = {Tancik, Matthew and Srinivasan, Pratul P. and
               Mildenhall, Ben and Fridovich-Keil, Sara and
               Raghavan, Nithin and Singhal, Utkarsh and
               Ramamoorthi, Ravi and Barron, Jonathan T. and Ng, Ren},
  title     = {Fourier Features Let Networks Learn High Frequency
               Functions in Low Dimensional Domains},
  booktitle = {Advances in Neural Information Processing Systems},
  volume    = {33},
  pages     = {7537--7547},
  year      = {2020}
}

@article{wang2026omnifacerig,
  author  = {Wang, Chao and Ma, Guangyao and Doublestein, John and
             Chen, Junming and Lin, Yiming and Su, Zhaoen and Luo, Xiaomin
             and Cheng, Shiyang and Shen, Jie and Roble, Doug and
             Wang, Dilin and Li, Yilei and Ranjan, Rakesh},
  title   = {{OmniFaceRig}: Fully Automatic Inner-Mouth-Aware Face Rigging
             Across Diverse 3D Character Topologies},
  journal = {arXiv preprint arXiv:2606.08043},
  year    = {2026},
  url     = {https://arxiv.org/abs/2606.08043}
}

@inproceedings{wang2026waft,
  author    = {Wang, Yihan and Deng, Jia},
  title     = {{WAFT}: Warping-Alone Field Transforms for Optical Flow},
  booktitle = {International Conference on Learning Representations
               (ICLR)},
  year      = {2026},
  url       = {https://openreview.net/forum?id=HTqGE0KcuF}
}

@article{lewis2010direct,
  title={Direct manipulation blendshapes},
  author={Lewis, John P and Anjyo, Ken-ichi},
  journal={IEEE Computer Graphics and Applications},
  volume={30},
  number={4},
  pages={42--50},
  year={2010},
  publisher={IEEE}
}

@article{hou2024neutral,
  title={Neutral Facial Rigging from Limited Spatiotemporal Meshes},
  author={Hou, Jing and Weng, Dongdong and Zhao, Zhihe and Li, Ying and Zhou, Jixiang},
  journal={Electronics},
  volume={13},
  number={13},
  pages={2445},
  year={2024},
  publisher={MDPI}
}

@article{li2010example,
  title={Example-based facial rigging},
  author={Li, Hao and Weise, Thibaut and Pauly, Mark},
  journal={Acm transactions on graphics (tog)},
  volume={29},
  number={4},
  pages={1--6},
  year={2010},
  publisher={ACM New York, NY, USA}
}

@article{xu2014controllable,
  title={Controllable high-fidelity facial performance transfer},
  author={Xu, Feng and Chai, Jinxiang and Liu, Yilong and Tong, Xin},
  journal={ACM Transactions on Graphics (TOG)},
  volume={33},
  number={4},
  pages={1--11},
  year={2014},
  publisher={ACM New York, NY, USA}
}

@inproceedings{thies2016face2face,
  title={Face2face: Real-time face capture and reenactment of rgb videos},
  author={Thies, Justus and Zollhofer, Michael and Stamminger, Marc and Theobalt, Christian and Nie{\ss}ner, Matthias},
  booktitle={Proceedings of the IEEE conference on computer vision and pattern recognition},
  pages={2387--2395},
  year={2016}
}

@inproceedings{richard2021meshtalk,
  title={Meshtalk: 3d face animation from speech using cross-modality disentanglement},
  author={Richard, Alexander and Zollh{\"o}fer, Michael and Wen, Yandong and De la Torre, Fernando and Sheikh, Yaser},
  booktitle={2021 IEEE/CVF International Conference on Computer Vision (ICCV)},
  pages={1153--1162},
  year={2021},
  organization={IEEE}
}

@article{lewis2014practice,
  title={Practice and theory of blendshape facial models.},
  author={Lewis, John P and Anjyo, Ken and Rhee, Taehyun and Zhang, Mengjie and Pighin, Frederic H and Deng, Zhigang},
  journal={Eurographics (State of the Art Reports)},
  volume={1},
  number={8},
  pages={2},
  year={2014}
}

@inproceedings{wang2023versatile,
  title={Versatile face animator: Driving arbitrary 3D facial avatar in RGBD space},
  author={Wang, Haoyu and Wu, Haozhe and Xing, Junliang and Jia, Jia},
  booktitle={Proceedings of the 31st ACM international conference on multimedia},
  pages={7776--7784},
  year={2023}
}

@inproceedings{carrigan2020expression,
  title={Expression Packing: As-Few-As-Possible Training Expressions for Blendshape Transfer},
  author={Carrigan, Emma and Zell, Eduard and Guiard, C{\'e}dric and McDonnell, Rachel},
  booktitle={Computer Graphics Forum},
  volume={39},
  number={2},
  pages={219--233},
  year={2020},
  organization={Wiley Online Library}
}

@inproceedings{onizuka2019landmark,
  title={Landmark-guided deformation transfer of template facial expressions for automatic generation of avatar blendshapes},
  author={Onizuka, Hayato and Thomas, Diego and Uchiyama, Hideaki and Taniguchi, Rin-ichiro},
  booktitle={2019 IEEE/CVF International Conference on Computer Vision Workshop (ICCVW)},
  pages={2100--2108},
  year={2019},
  organization={IEEE}
}

@inproceedings{dutreve2010easy,
  title={Easy rigging of face by automatic registration and transfer of skinning parameters},
  author={Dutreve, Ludovic and Meyer, Alexandre and Orvalho, Veronica and Bouakaz, Saida},
  booktitle={International Conference on Computer Vision and Graphics},
  pages={333--341},
  year={2010},
  organization={Springer}
}
